\documentclass[review]{elsarticle}

\usepackage{lineno,hyperref}
\modulolinenumbers[5]

\journal{IEEE System Journal (Accepted)}

\usepackage{amsmath}
\usepackage{graphics}
\usepackage{graphicx}
\usepackage{amsmath}
\usepackage{times}
\usepackage{mathrsfs}
\usepackage{array}
\usepackage{amsmath, latexsym, amsfonts, amssymb}
\usepackage[table]{xcolor}
\usepackage{multirow}
\usepackage{textcomp}
\usepackage{booktabs}

\ifodd 1

\else
\fi

\begin{document}

\begin{frontmatter}

\title{Urban Space Insights Extraction using Acoustic Histogram Information}

\author{Nipun Wijerathne}
\ead{hnipun@gmail.com}
\author{Billy Pik Lik Lau\corref{mycorrespondingauthor}}
\ead{billy\_lau@mymail.sutd.edu.sg }
\cortext[mycorrespondingauthor]{Corresponding author}
\author{Benny Kai Kiat Ng}
\ead{benny\_ng@sutd.edu.sg}
\author{Chau Yuen}
\ead{yuenchau@sutd.edu.sg}
\address{Singapore University of Technology and Design (SUTD), 8 Somapah Rd, Singapore 487372}

\begin{abstract}
Urban data mining can be identified as a highly potential area that can enhance the smart city services towards better sustainable development especially in the urban residential activity tracking. 
While existing human activity tracking systems have demonstrated the capability to unveil the hidden aspects of citizens' behavior, they often come with a high implementation cost and require a large communication bandwidth.
In this paper, we study the implementation of low-cost analogue sound sensors to detect outdoor activities and estimate the raining period in an urban residential area. 
The analogue sound sensors are transmitted to the cloud every 5 minutes in histogram format, which consists of sound data sampled every 100ms (10Hz).
We then use wavelet transformation (WT) and principal component analysis (PCA) to generate a more robust and consistent feature set from the histogram. 
After that, we performed unsupervised clustering and attempt to understand the individual characteristics of each cluster to identify outdoor residential activities.
In addition, on-site validation has been conducted to show the effectiveness of our approach. 
\end{abstract}

\begin{keyword}
 Smart City, Activity Detection, Sound Histogram, Sound Sensor, Clustering. 
\end{keyword}

\end{frontmatter}


\section{Introduction}
Smart cities allow integrated systems to monitor the human and environmental behaviors in real-time have been looked upon as a potential solution to fulfill the requirements of modern urbanization.
These statistics related to behaviors and knowledge are then used to make more effective decisions to further improve the urban design and living standards of people. 
With the recent enhancements of wireless networks, large-scale systems for various sectors can be realized easily, which resides an enormous amount of spatial-temporal data that can be harvested.
These data contain many interesting aspects of day-to-day utilization of an urban space or life of citizens\cite{Katakis2015Mining,Lau2018Sensor,Lau2019survey,jayasinghe2019feature,Wijerathne2018Towards,Lau2016Spatial}.
With the advancement of machine learning and data mining technique, different platforms have been developed to cater different usages of monitoring in a smart city.
Various aspects have to be taken into consideration when deploying a smart city project, such as deployment and maintenance cost, system and data scale, sensor coverage and resolution, user privacy, data storage, etc.
Take all these factors into design and implementation consideration, urban monitoring in a smart city is extremely challenging. 
The increasing population density has driven more efficient real estate planning and at the same time needs to balance out aspects of liveable space in a dense area. 
Hence, the urban designer needs to understand the residential behavior of the existing urban area.
Common technique such as outdoor activity detection is often used and it related to occupancy detection (can be indoor also).
Such activity detection can be identified using sensors such as thermal imaging~\cite{Tyndall2016Occupancy}, camera~\cite{Liu2017Human,Yu2019Towards}, motion\cite{Lau2016Spatial,Hao2009Multiple}, mobile devices\cite{Booranawong2019Adaptive,Zhou2020Understanding,Li2019Experimental}, and sound sensor~\cite{Bieber2011Hearing,Longo2017Crowd,Fairbrass2019CityNetDeep}.
A lot of the aforementioned techniques require high computation cost or communication bandwidth, which may not be suitable for continuous smart city monitoring.
Not to mention, privacy intrusion is also another concern among researchers \cite{kitchin2016getting, bartoli2011security}.

In this paper, we focus on discovering human outdoor activity detection techniques and patterns in an urban point of interest (PoI) using low-cost analogue sound sensors.
We want to show that using a low-cost analogue sensor is able to capture human outdoor activity while maintaining the sustainability of a sensor node and reducing communication overhead. 
However, the challenge in using the sound sensor is that the outdoor environment tends to have a larger variance in ambient noise and hence it is harder to distinguish outdoor activity as well as background noise.
Also, another huddle in collecting sound data is the sampling rate of the sensor, which may consume more power as well as constant communication between the gateway to transmit data. 
Thus, our motivation is to design a system model that process the histogram data from sound sensors to extract insights such as human activity and raining period of an urban PoI through compressed information such as sound histograms.

We leverage the idea of edge computation to achieve self-sustainability of the sensor node (using energy harvesting) at the same time to minimize the communication bandwidth between sensor nodes and gateway.
The analogue sound sensor capture sound intensity collected from the surrounding environment, which can be composed of people chatting, children playing, pass-by vehicles, rain, etc.
The sound data collected are complied at the sensor node into a sound histogram and uploaded to the cloud database every $5$-minute for off-line processing. 
We first preprocess the sound histogram data using wavelet transformation (WT) and principal component analysis (PCA) to extract a more robust feature sets.
WT has been extensively studied in area of speech recognition~\cite{Rekik2012Speech}, outliers detection~\cite{Bilen2002Wavelet}, and image processing~\cite{Kolekar2018introduction}. 
Based on the literature, it seems like a natural selection to use WT for developing a richer feature space for sound histogram by allowing changes in the temporal aspect.
Next, we use PCA to convert the transformed sound histogram into a set of uncorrelated values, which can be further utilized for insights extraction. 
PCA is studied in~\cite{Lever2017Points,han2011data,Abegaz2018Principals} and shows effective in generating uncorrelated values while reducing data dimension for easier visualization.
Here, we divide the insights extraction process into two parts, which are raining period estimation and outdoor activity detection.
The raining period is an example of a common event impact multiple sensors when compared to outdoor activity, where the event only can be detected by a single sensor.
In addition, the raining period estimation is important in outdoor activity detection as raining day affects outdoor public space utilization. 
Similar principle can be extended to detect large area events such as explosion or party of a large group of participants, where multiple sensors are impacted simultaneously.
To detect outdoor activity, we utilize the clustering method to group sound samples with high similarity and attempt to understand background sound according to the time of the day and day of the week. 
Through the identified background sound, the different period has been defined and chi-square statistic is used for distinguishing the background sound and activity. 
The key challenge here is to classify the human outdoor activity and background sound accurately with a simple and effective method. 
Ground-truth data is also collected for validating the system model and show its effectiveness in a real-world scenario.

The contributions of our work in this paper are three-fold:
\begin{itemize}
	\item We introduce the concept of compressing the sound samples of an urban environment as a histogram leveraging edge-computation techniques.
	\item We propose a system with preprocessing (WT and PCA) as well as clustering techniques to obtain useful feature set for identifying background noise and human outdoor activity. We show that the proposed system model is capable to achieve accuracy of $85.8\%$ distinguishing background sound and human outdoor activity.
	\item Using multiple sensor nodes, we can estimate events that cover large areas, where we use the raining period estimation as an example, to show that using our proposed method we can get a better estimation than using a resistive water droplet sensor.
\end{itemize}
\vspace{-0.15cm}

The remaining of this paper is organized as follows:
Section~\ref{sec_sysModel} presents the system for processing the sound data as well as the details of the sound histogram data. 
Next in Section~\ref{pre}, we discuss the details used for preprocessing the sound histogram and methods for clustering processed sound samples.
Section~\ref{OD} describes the raining period estimation and human outdoor activity detection method, followed by Section~\ref{experi}, which validates our outdoor detection method against ground-truth.
Lastly, we conclude our work in Section~\ref{concl}.

\section{System Model}
\label{sec_sysModel}

The system model consists of three different stages, which are (1) data collection phase, (2) data preprocessing and features generation phase, as well as (3) insight extraction phase.
The overall system model is shown in Fig.~\ref{fig10_validation}.

\begin{figure}[h]
	\centering 
	\includegraphics[width=0.52\textwidth]{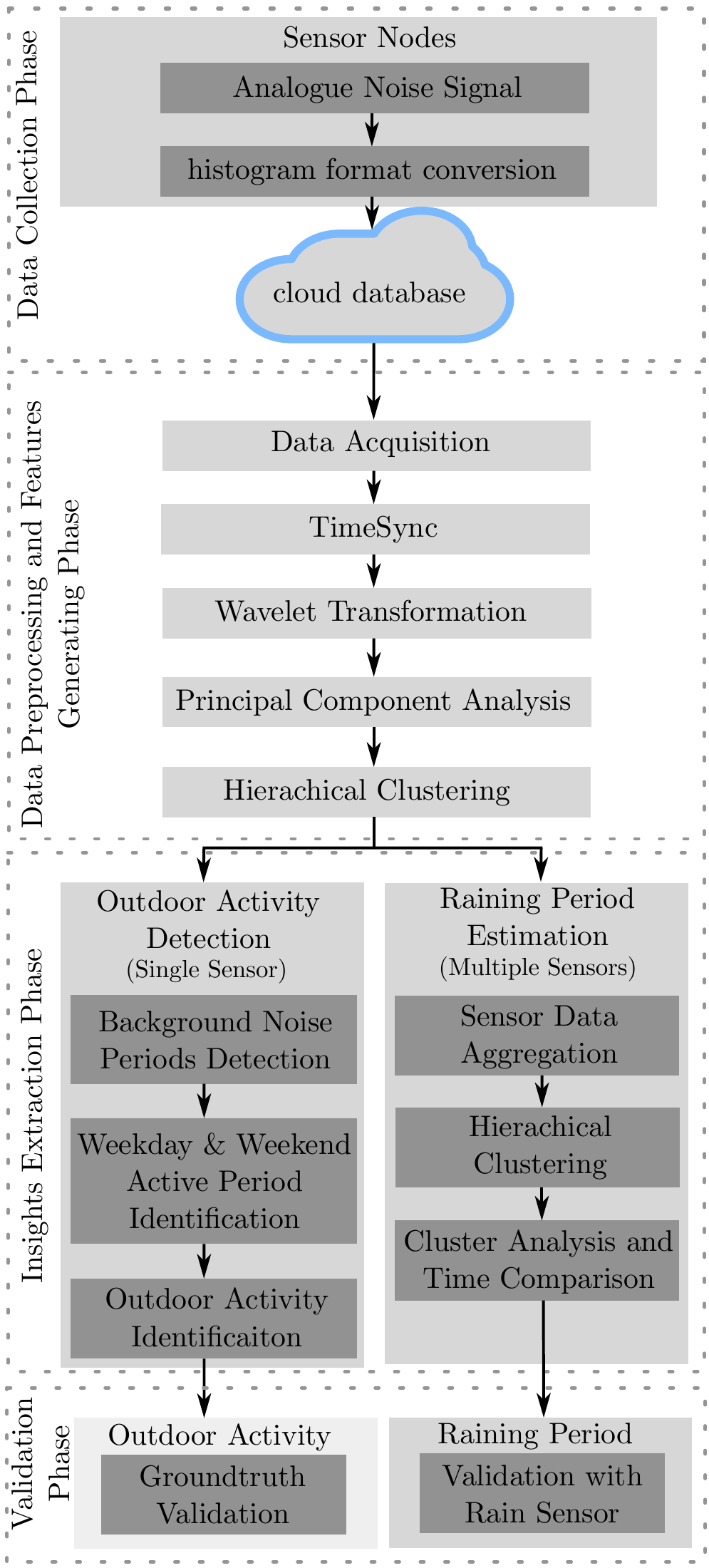} 
	\caption{Sound Data Processing System Model}
	\label{fig1_systemModel}
\end{figure}

The first stage involves data collection using sensor nodes in an urban residential area, which is first introduced by ~\cite{Lau2018Sensor,Lau2016Spatial}. 
The following describes the methodology of generating sound histograms from sound samples:
we use an analogue sound sensor (dfrobot Analog Sound Sensor SKU: DFR0034) and it is capable of detecting the ambient sound intensity using LM386 amplifier and electret microphone.
The output of the sensor (noise intensity level) is an analog value, and we use a $10$-bit ADC to quantize it into $1024$ levels.
It covers the range of $0$ to $1023$ with a higher value that indicates higher noise intensity.
The sound sensor data is then sampled in the rate of every $100$ms ($10$Hz sampling frequency).
For every $5$ minutes, we would have $3000$ sound samples collected and this will be fitted into histogram $h(m)$.
\vspace{-0.15cm}
\begin{equation}
h(m) = \lbrack s_{1} ,..., s_{m}\rbrack
\vspace{-0.15cm}
\end{equation}
where $\bf h$ is denote as the vector notation. $s_{m}$ is the $m^{th}$ bin value and $m\in {\{1,...,5\}}$.
These $3000$ sound samples are divided into $5$ bins of a histogram based on the value ranges.
The rationality of choosing $5$ bins of the histogram is that the majority of idling time (without any human activity), noise intensity has a lower range of value. 
Therefore, any higher value denotes chances of activity presence regardless of value number (i.e. value of $50$ vs $333$), which varies on noise source distance.
Also, through the number of bin values, we can observe the duration of human activity.
The following Table~\ref{tbl_soundRange} show the range of quantifying the $0$ to $1023$ noise value:

\begin{table}[h]
	\caption{Sound Sensor Intensity Ranges}
	\label{tbl_soundRange}
	\begin{center}
		\begin{tabular}{|c||c||c||c||c|}
			\hline
			Range 1 & Range 2 & Range 3 & Range 4 & Range 5\\
			\hline
			$0\le i\le6$ & $6< i\le10$ & $10< i\le20$ & $20< i\le50$ & $i > 50$
			\\
			\hline
		\end{tabular}
	\end{center}
\end{table}

To further elaborate the consolidation process of sound samples, an example as follows is presented:
Using $1$ second sampling time as a baseline, we have a $10$ sound samples collected, such as $[11,2,7,33,55,80,28,7,9,13]$.
Based on the collected samples, we consolidate them into a histogram of $[1,3,2,2,2]$.
The rationality of choosing this method is to help to cope with the communication bandwidth requirements and reduces the communication overhead.
Whenever the histogram has been consolidated, it will be uploaded to the cloud database via a wireless sensor network gateway every $5$ minutes. 

The second processing stage involves preprocessing techniques previously introduced in~\cite{Lau2018Sensor,Lau2016Spatial} such as data acquisition and TimeSync.
After performing cleaning up on the data, cleaned up histogram data are further processed to generate more features for insights extraction.
In this paper, we focus on two kinds of insights extraction from the sound sensor`s histograms data format, which are (1) human outdoor activity detection and (2) raining period estimation.
However, the former will be given more focus as it is the main purpose of this paper - to detect human outdoor activity in an urban area using analogue sound sensor. 
The latter will only be briefly discussed as we discover that it is possible to estimate the raining period using the histogram sound data aggregated from multiple sensors.
Further details of both insights extraction methods will be elaborated in the upcoming session.

\section{Sound Histograms Preprocessing and Clustering Phase} 
\label{pre}
In this section, we discuss the in-depth techniques used to process the sound histogram.
First, we explain the WT approach applied to transform the sound histograms into feature spaces. 
Subsequently, we utilize the PCA method to generate richer feature sets from WT applied sound histograms.

\subsection{Wavelet Transformation} \label{wave}
WT based feature extraction method is used to develop the feature space for sound histograms.
Wavelet coefficients for a sound histogram are computed using a series of dilation and translation of the mother wavelet~\cite{Leavey2003introduction,Sarkar1998tutorial}. We can represent the $m^{th}$ bin of a sound histogram as follows:
\begin{equation}
h(m) = \sum_{j}\sum_{k}d_{j,k}\psi_{j,k}(m), 
\label{eq1}
\end{equation}
where $\psi_{j,k}(m)$ is the scaled and dilated mother wavelet function, the $d_{j,k}$ is the wavelet coefficient that represents how much translated and dilated mother wavelet describes the given sound histogram $h(m)$, and $ j,k \in \mathbb{Z}$.
from there, we can formulate the following equation:
\begin{equation}
d_{j,k}= \sum_{m}h(m)\psi_{j,k}(m)= <h, \psi_{j,k}>,
\end{equation}
since wavelets are orthogonal to each other.

Wavelets are simply mathematical functions that exhibit only localized oscillations, which decay quickly.
This allows us to study local undulations of the sound histograms sound sample. 
For example, if the given sound histogram $h(m)$ has a discontinuity, then it will only influence the $\psi_{j,k}(m)$ that is in the vicinity.
Only those coefficients $\{d_{j,k} \}_{j,k \in \mathbb{Z}}$ whose associated wavelet $\psi_{j,k}(m)$ overlaps the discontinuity will be influenced.
Moreover, it should be noted that the bases of other common transformations are affected by the discontinuity, regardless of where it is located.

The localized behavior of the WT is important in generating the feature set for the histograms sound data. 
Thus, we use the \textit{Haar} basis function as the mother wavelet due to its discontinuous nature. 
It can be used to analyze histograms effectively and efficiently at various resolutions and used to get the approximation coefficients and detail coefficients at various levels as shown in~\cite{Leavey2003introduction}.
Besides, it is also served as a low-pass filter and a high-pass filter simultaneously.
The \textit{Haar} basis function can de denoted as follows:
\begin{equation}
\psi_{j,k}(x) = \begin{cases}
\enspace 1  & \text{if } x \in [0,1/2) \\
-1  & \text{if } x \in [1/2,1) \\
\enspace 0  & \text{otherwise } 
\end{cases}.
\end{equation}
In our system model, $5$ minutes sound histogram sound samples will be represented in \textit{Haar} basis function, and those wavelet coefficients $\{d_{j,k} \}_{j,k \in \mathbb{Z}}$ are used as the feature vector. 
Subsequently, discrete \textit{Haar} wavelet coefficients can be calculated as 
\begin{equation}
\bf w=B^{-1}h,
\end{equation}
where $ \bf B $ is the wavelet basis matrix and $\bf w$ is the wavelet coefficient vector for the sound histogram. 
Being restricted to dyadic sequences, traditional \textit{Haar} wavelet construction is not sufficient.
The reason due to the number of sound histogram bins is non-dyadic. 
Therefore, the intuition behind this method is to use a non-dyadic \textit{Haar} construction, which is more accurate than the zero paddings.

Let us denote the elements at level $i$, interval $j$ by $ T_{i,j} $ and $ |T_{i,j} |  = n_{i,j}$. 
Here, the elements can be consists of the following: $ T_{0,1} = \{s_{1}, s_{2}, s_{3}, s_{4}, s_{5}\} $, $  T_{1,1}= \{s_{1}, s_{2}, s_{3}\} $, $  T_{1,2}=\{s_{4}, s_{5}\} $, $  T_{2,1} = \{s_{1}, s_{2}\} $, and $ T_{2,2} = \{s_{3}\} $.
The length of intervals being $ n_{0,1}= 5, n_{1,1}= 3, n_{1,2}= 2, n_{2,1}= 2, n_{2,2}= 1$. 
We can represent the elements of $\bf B $ that corresponds to detailed coefficients as follows:
\begin{equation}\label{eqn16}
B_{l,m}= \begin{cases}
\enspace \displaystyle(\sqrt{n} \frac{1}{2}\sqrt{\displaystyle\frac{n_{a}+n_{b}}{n_{a}n_{b}}})\displaystyle\frac{2n_{b}}{n_{a}n_{b}}  & \text{if } s_{m} \in T_{i+1,k} \\\\
-(\sqrt{n} \displaystyle\frac{1}{2}\sqrt{\displaystyle\frac{n_{a}+n_{b}}{n_{a}n_{b}}})\displaystyle\frac{2n_{a}}{n_{a}n_{b}}  & \text{if } s_{m} \in T_{i+1,k} \\\\
\enspace 0  & \text{otherwise } 
\end{cases}.
\end{equation}
where the elements of $ \bf B $ are represented by $ B_{l,m} $ ($ B_{1} $ to $ B_{4} $) and $ n_{a}= n_{i+1,k} $ , $ n_{b}= n_{i+1,k+1} $. $ B_{0} $ \textit{i.e.}, the elements corresponds to average coefficients can be given as $ B_{0} = \{1, 1, 1, 1, 1\} $.

Elements of $\bf B$ in the Equation~\ref{eqn16} can be calculated using the following matrix:
\begin{equation}
\bf B = \begin{pmatrix} B_{0} \\ 
B_{1} \\ 
B_{2} \\ 
B_{3} \\ 
B_{4} \\   
\end{pmatrix} = \begin{pmatrix} 1 & 0.82 & 0.91 & 1.58 & 0.00 \\ 
1 & 0.82 & 0.91 & -1.58 & 0.00 \\ 
1 & 0.82 & -1.82 & 0.00 & 0.00 \\ 
1 & -1.22 & 0.00 & 0.00 & 1.58 \\ 
1 & -1.22 & 0.00 & 0.00 & -1.58 \\   
\end{pmatrix}
\end{equation}

This concludes our work on the wavelet transforms for the sound histogram. 
Subsequently, we will perform a dimension reduction technique for the wavelet transformed histograms sound samples using PCA.

\subsection{Principle Component Analysis} 
\label{prin}
\begin{figure}[h]
	\centering {\includegraphics[width=0.95\textwidth]{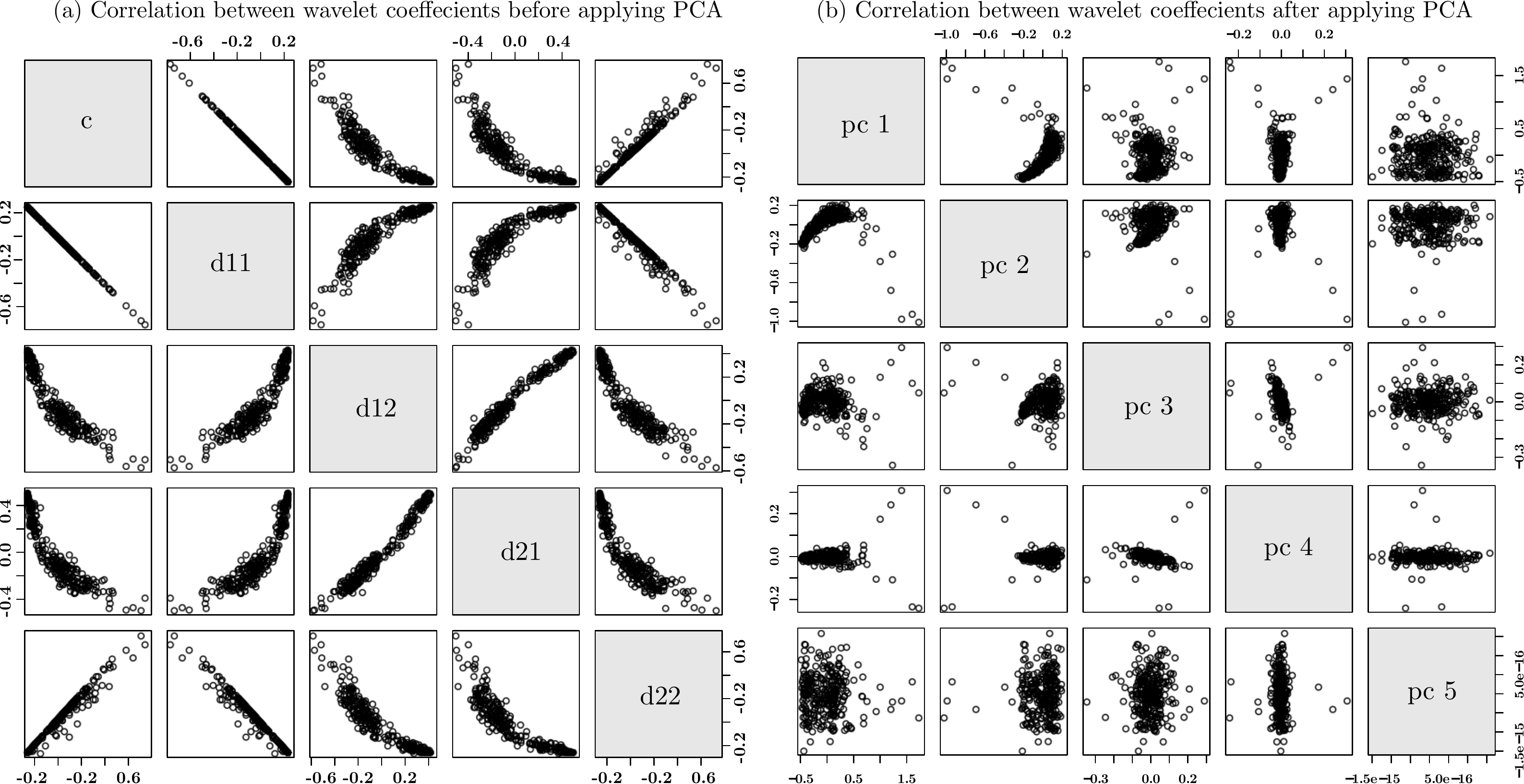}} 
	\caption{Comparing the correlation coefficient between wavelet coefficients (a) before PCA (b) after PCA. }
	\label{fig2_correlation}
\end{figure}

Histogram sound samples provide compressed information about ambient sound intensity but also induces a high correlation between each bin. 
This causes instability in distance metric calculation when performing clustering.
Moreover, we can simply prove that this high correlation is not affected by the WT by given the fact that it is a linear transformation (by using the fact $W[X]=aX+b$ then, $ W[\bar{X}] = a\bar{X}+b$). 
The result of such is displayed in Fig.~\ref{fig2_correlation}.

Fig.~\ref{fig2_correlation}(a) illustrates the correlation between $ d_{j,k} $, which is calculated for a given sensor node.
It is observed that applying PCA, the high correlation between data has been removed. 
Note that, the column and row represent the coefficients and the scattered plot represents the correlation between two different coefficients.
Thus, it is not hard to see that there is a high correlation between $ d_{j,k} $. 
We transform the sound histogram into an uncorrelated space using PCA to remove the high correlation in between.
Consequently, PCA provides an optimal solution for the following optimization problem:

\begin{equation}
\begin{cases}
\text{max} & e_{i}^T\Sigma e_{i}\\
\text{subject to} & e_{i}^Te_{i} = 1 \\
& e_{j}^T\Sigma e_{i} = 0,  \text{  for } j< i,
\end{cases}
\end{equation}
where $\Sigma$ is the covariance matrix of calculated $ d_{j,k} $ for sound histogram (for one day of a given sensor node).
$e_{i}$,$e_{j}$ are the eigenvectors associated with $\Sigma$ in the optimal solution, which only satisfies the Karush Kuhn Tucker (KKT) optimality conditions. 
We denote the sound histograms for a one day ($288$ represents data collected for $5$-minute windows over $24$ hours) single sensor node as $ \bf H =\{w_{1},w_{2},\ldots, w_{288}\}$ and $\bf w$ is the wavelet coefficient vector for each sound histogram.
Afterward, the projection of $\bf H$ onto \textit{PC}s can be represented as follows:
\begin{equation}
\bf Z = \begin{pmatrix} e_{1} \\ 
. \\ 
. \\ 
. \\ 
e_{5} \\   
\end{pmatrix} H
\end{equation}
where $ \bf Z $ is the projection of $\bf H$. 
Fig.~\ref{fig2_correlation} shows the transformation of sound histograms into uncorrelated space after applying PCA.

Let $\bf U = \{e_{1},...,e_{5}\}$ then the reconstructed sound histograms $\bf \hat H$ from \textit{PC}s can be rewritten as shown below:
\begin{equation}\label{pca}
\bf \hat H = \bf UZ
\end{equation}
Mathematically $ \bf \hat H = H $ when we reconstruct the sound samples by including all the PCs.
Using this method, a truncated reconstruction with only a selected subset of PCs would yield some components being discarded. 
This can be used to remove redundant information and hence a cleaner data.
This method has been used in~\cite{Zhang2009PCA, Jagadesh2016novel}, where redundant information is removed while preserving a relatively low variation of data. 

\begin{figure}[h]
	\centering 
	\includegraphics[width=0.95\textwidth]{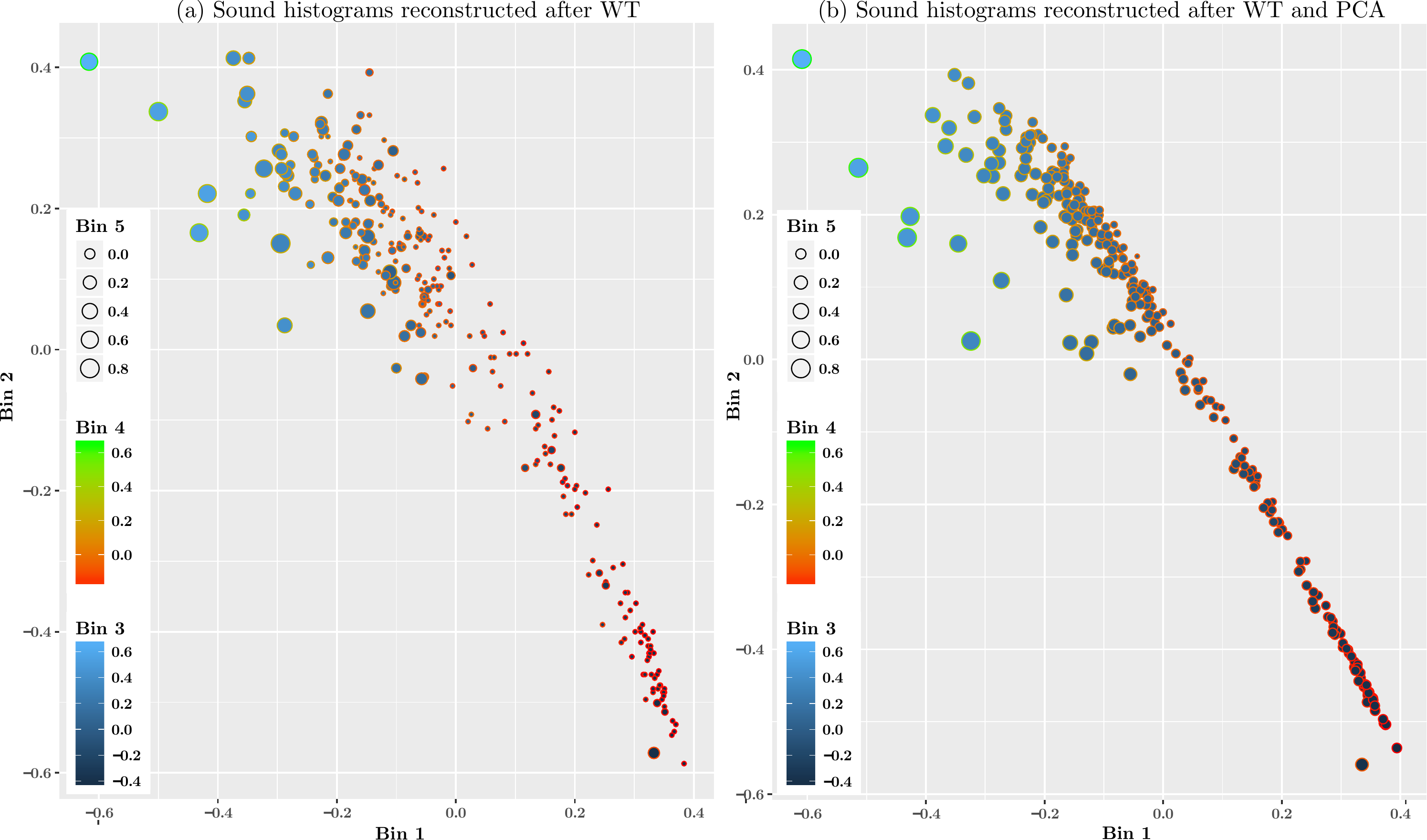} 
	\caption{Comparison of sound data using different data processing techniques. The first two dimension of the histograms are shown in $x$ and $y$ axes, while magnitude of the $3$rd dimension is given by the area of each circle. The $4$th dimension is shown by the color intensity level of the perimeter of each circle and the $5$th dimension is presented by the color intensity level of the area for each circle.} 
	\label{fig3_denoise}
\end{figure}

As shown in~\cite{Cattell1966Scree,Carreira-Perpinan1997review}, there is a few formal ways of selecting the optimal subset of PCs $\{pc_{1},...,pc_{k}\}$.
We select $k$ by minimizing the projection error between $\bf \hat H$ and $ \bf H$ given that $k<5$. 
Fig.~\ref{fig3_denoise} (a) and Fig.~\ref{fig3_denoise} (b) illustrate the $\bf H$ and  $\bf \hat H$ respectively.
The reconstructed sound histograms $\hat{\textbf{H}}$ are constructed using a subset of principal components from $\textbf{Z}$.
We select $k=4$ (4 principal components) and the simple illustration of such can be found in Fig.~\ref{fig3_denoise}.
It can be observed that $\bf \hat H$ has successfully removed the low variational components across all five sound histogram bins when compared with the $\bf H$.
By removing the low variational components, it also removes the redundant information inside the sound histograms.
Therefore, $\bf \hat H$ can guarantee a more robust and accurate feature representation of sound histograms.

\begin{figure}[h]
	\centering \includegraphics[width=0.86\textwidth]{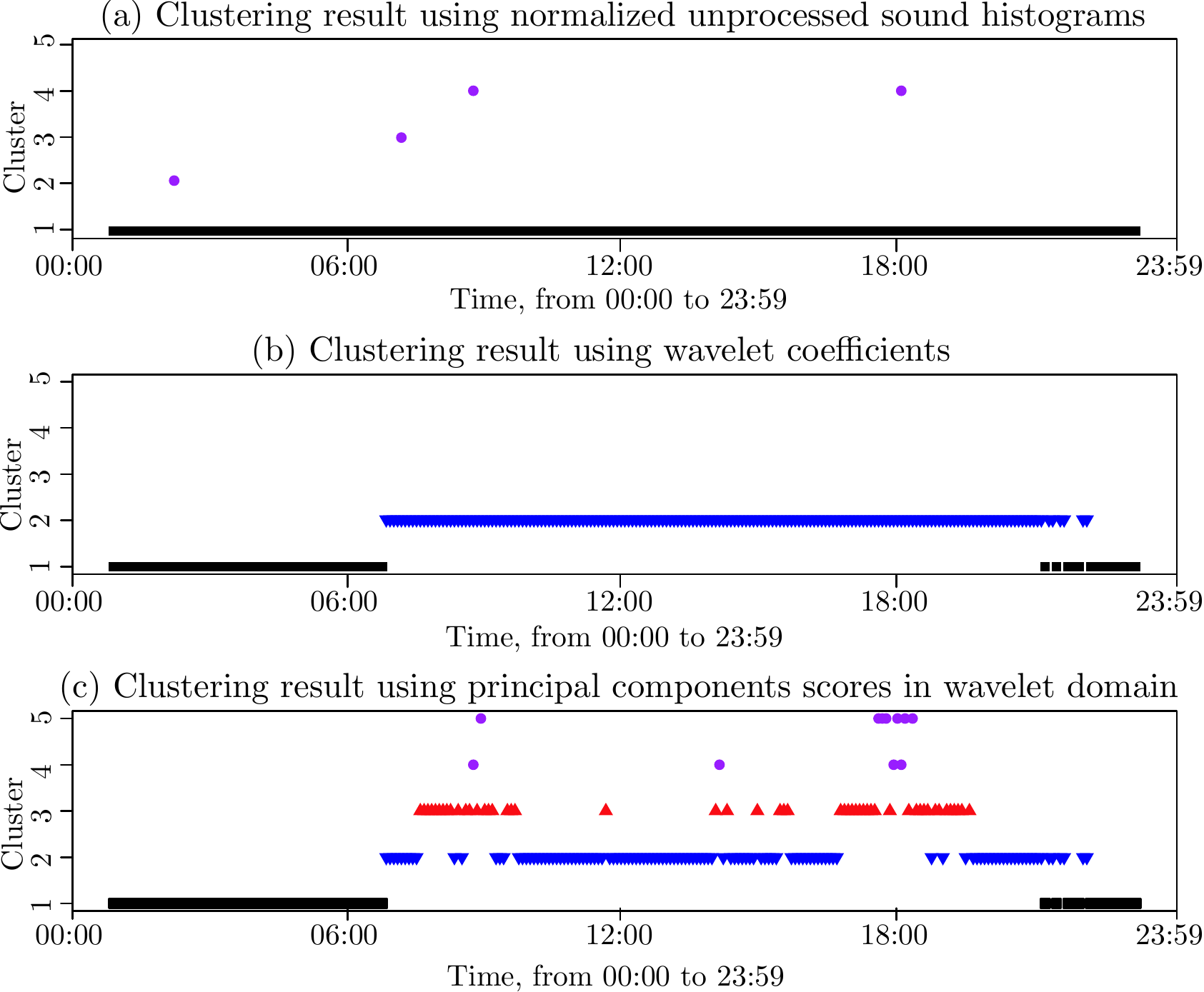}
	\caption{Clustering result using different preprocessing techniques: (a) using only normalized unprocessed sound histograms. (b) using wavelet coefficients. (c) using principal components scores in wavelet domain. }	 
	\label{fig4_clusterCompare}
\end{figure}

To validate the efficiency of the aforementioned data processing technique, we perform a simple hierarchical clustering method to show the difference between WT and PCA in terms of feature extraction based on sound histograms' similarity.
Fig.~\ref{fig4_clusterCompare} illustrates how the WT and the PCA enhance feature extraction based on the similarity of data.

Note that, the main focus here is to show the clustering result performed on different processing techniques rather than the technical detail of clustering (will be discussed in the next section).
In addition, the markers presented in the figure do not yield any similarity between data processing despite the markers are using the same colors.
It only represents the similarity corresponding to the exact data sample throughout the day, where it forms a different clustering group.
Once again, this does not show the activities being detected but rather displays the group of similar processed sound samples.
Fig.~\ref{fig4_clusterCompare}(a) shows that most data points are quite similar to each other for most of the time. 
Despite $4$ clusters being generated, it does not yield any meaningful insight. 
In Fig.~\ref{fig4_clusterCompare}(b), we observe there is a different cluster that appears in the morning to the evening period. 
However, this still does not show the variation between morning and evening. 
Lastly, after performing PCA and WT as shown in Fig.~\ref{fig4_clusterCompare}(c), we notice there are variations in the morning, afternoon, and evening. 
But this requires further investigation, which we will discuss in the subsequent sub-section. 
Here, we show that by applying WT and PCA techniques, it is possible to extract more variation in the histogram sound samples for further insights' extraction.

\subsection{Sound Histograms Clustering} 
\label{clustering}

In this sub-section, we focus on the clustering technique for processed sound data using the aforementioned techniques.
First, we calculate the average value for each $5$-minute sound histogram, and we noticed there is a latent hierarchical structure in sound histograms despite variation of behavior for different days of the week (Weekday, Saturday, and Sunday).
Therefore, it is crucial to trench this hidden structure in order to establish a model for daily usage using the agglomerative hierarchical clustering method~\cite{Murtagh2012Algorithms}.

\begin{figure}[t]
	\centering
	\includegraphics[width=0.99\textwidth]{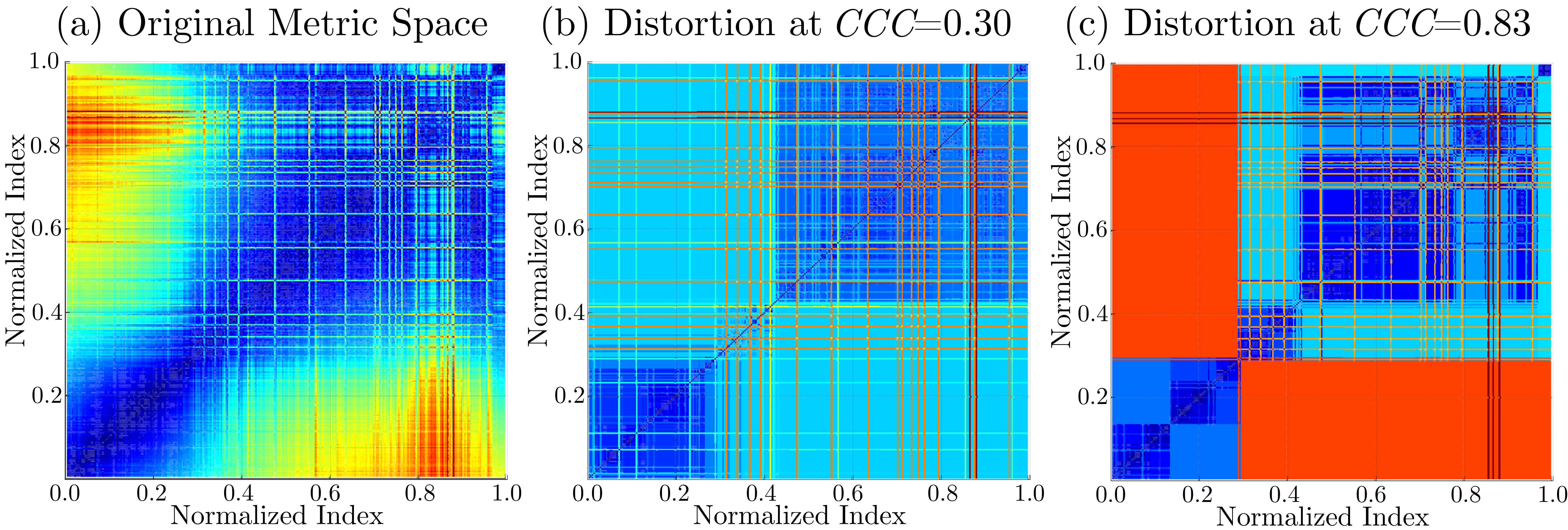}
	\caption{Comparison of distortion of metric space for sound histograms.}
	\label{fig5_CCC}
\end{figure}

Agglomerative hierarchical clustering method provides variation to the distance measure between two sound histograms samples, where the only measurement is needed to formulate the cluster.
This simplifies the clustering process for the PCA and WT results, while provide a visual clue to diagnose the clustering result.
In our study, we generate distortion based on the distance between two sound histogram samples using Euclidean distance.
We observe such behavior in sound histogram samples and it varies depending on different environmental conditions and time of the days.
Therefore, it is crucial to find the correct similarity measurements for sound histograms on each day for each individual sensor nodes, so that we can reduce the distortion of the metric spaces. 
We calculate the Cophenetic Correlation Coefficient (\textit{CCC}) as a goodness of fit static for the hierarchical dendrogram for each similarity measurement type. 
We study two different \textit{CCC} values as illustrated in the Fig.~\ref{fig5_CCC}. 

The $x$ and $y$ axes denote the normalized index of the sound histogram, which can be calculated through min-max normalization of sound histogram w.r.t the number of sound histograms samples collected in a day.
By visually observing the distortion of metric space, it is not hard to see that higher~\textit{CCC} value resembles the original metric space provides a more meaningful clustering outcome. 
Next, we choose the dendrogram that has the highest~\textit{CCC} value that shows the best clustering consistency as the foundation for clustering similar background noise.

\section{Detection Phase} \label{OD}
After defining the data preprocessing methods, we analyze the clustered data in order to detect events such as (1) raining period, (2) background noise, and (3) human outdoor activity.

\subsection{Raining Period Estimation}
\label{RPE}

Weather plays an important role in affecting human outdoor activity. 
Hence, it is critical to avoid including the raining period as part of human outdoor activity detection by estimating the raining period.
To better estimate the raining period, multiple sensors ($7$ sensor nodes that are co-located in the same playground in our case) within the same PoI are used. 
The rationality of using multiple sensor nodes is that the rain is the common event that will be captured by all the sensors in the vicinity at the same time-line.
To provide a comparison for raining period detection, resistive water droplet sensors are used for validation.
The resistive water droplet sensors output a value ranging from $0$ to $1023$, where lower value denote presence of water droplet while higher value denote the opposite.
Note that, the actual raining period detected by resistive water sensors can be shorter due to the remaining water droplet on the sensor surface.

\begin{figure}[h]
	\centering
	\includegraphics[width=0.85\textwidth]{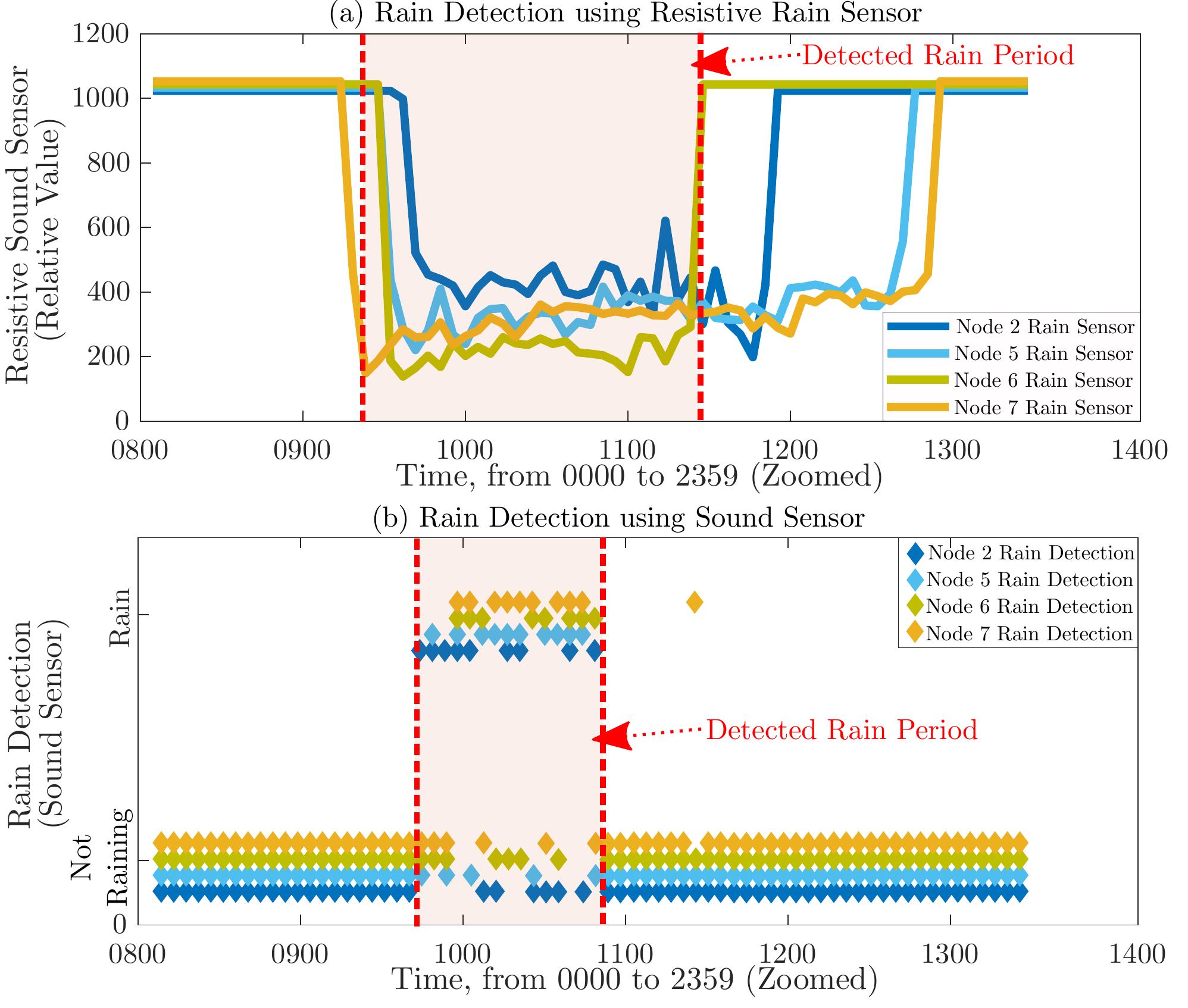}
	\caption{Comparing rain detection method using (a) rain sensor data (solid line) and (b) noise sensor data (diamond markers) based on the data collected on $28$th July $2016$. }
	\label{fig6_rainDetection}
\end{figure}

We use the clustering method for $7$ sensor nodes' sound histogram, and we observe that are cluster that resembles raining period compared to non-raining period.
To further validate the accuracy of the raining period, we investigate the resistive raining sensor against the estimated raining period as shown in Fig.~\ref{fig6_rainDetection}.

The solid line represents resistive rain value for each sensor and the diamond markers denote raining period detected using sound sensor.
We notice there is a slight delay in identifying the raining period when compared it to the resistive rain sensor. 
However, in term of actual raining period, sound sensor can estimate the time of rain stop more accurate than the resistive round sensor. 
To better separate the raining period detection and human outdoor activity, prolong period of high ambient noise intensity is one of the significant characteristic. 
Any other short impulse of noise periods (e.g. heavy trucks, car, thunder, train, etc.) would be discarded when compared to the actual raining period.
Thus, we show that the raining period estimation is possible using the clustering method given the interval of data samples is sufficient.

\subsection{Background Noise Detection}
\label{backg}

\begin{figure}
	\centering 
	\includegraphics[width=1.0\textwidth]{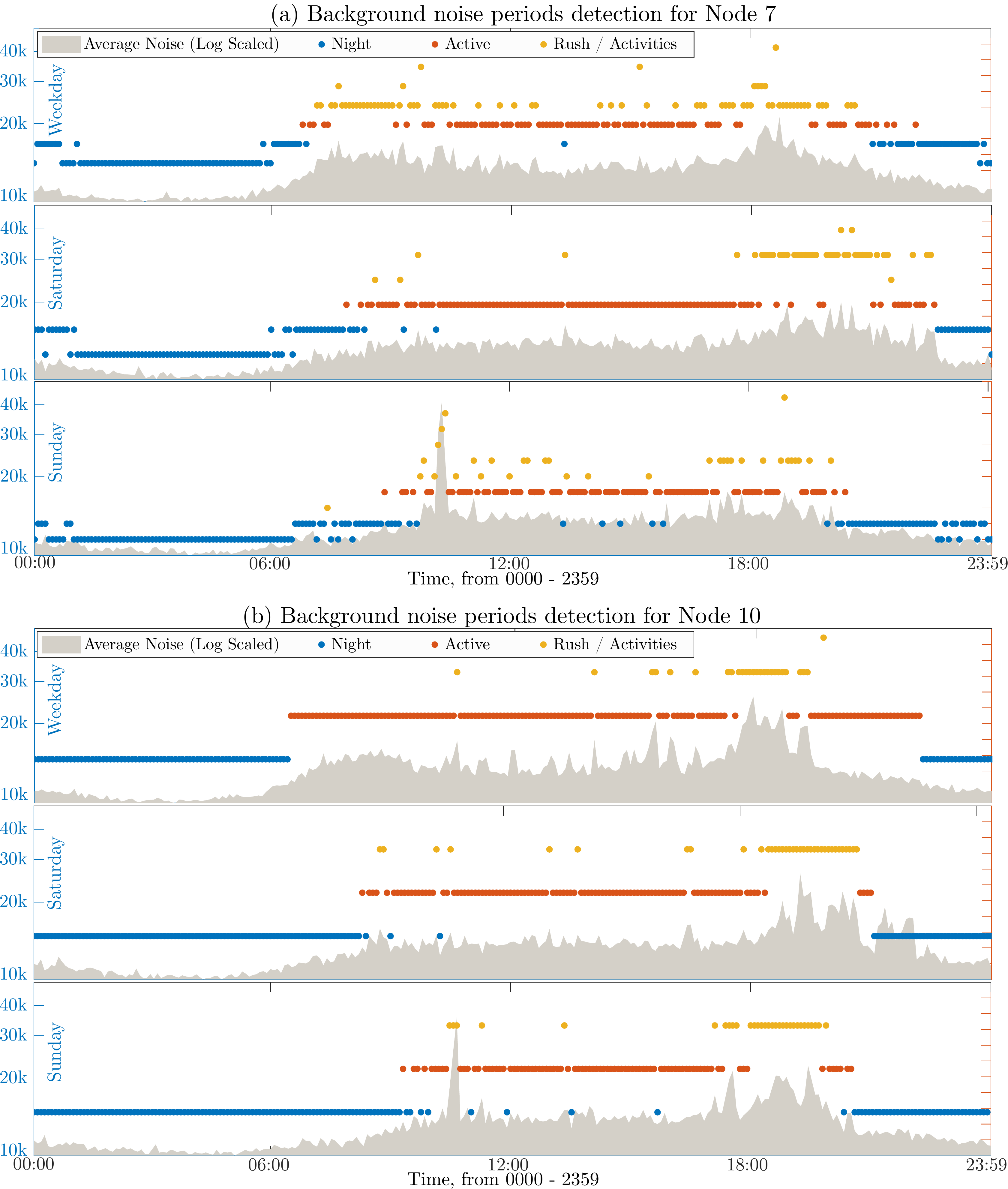} 
	\caption{Background noise periods detection for sensor Node $7$ and $10$. (Top: Weekday $2$nd August $2016$. Middle: Saturday $16$th July $2016$. Bottom: Sunday $10$th July $2016$.)}
	\label{fig7_backgroundAvgNoise}
\end{figure}

Before identifying the human outdoor activity, we need to understand the characteristic of background noise period for each sensor node at the given PoI.
First, we define the ambient background noise in our system model given the human activity is absent. 
The most straightforward approach to understand the background noise is to restrict any potential human activities for weeks to record a human activity free background noise, but this method would prevent the resident to utilize the PoI.
Thus, we attempt to understand the background noise based on an assumption that the human activity does not occur at the same time throughout the week and months.
Based on this assumption, we can detect the human activity using the outliers detection method (will be further elaborated in the next sub-section~\ref{OAD}).
In order to identify the background noise, we use the aforementioned clustering technique with the processed data (applying WT and PCA) for each day.
An example of the background noise periods throughout different day of the week can be found in Fig.~\ref{fig7_backgroundAvgNoise}.

By observing the general patterns of the background period, we divide it into the following categories in Table~\ref{tbl2_definition} to match their characteristic:

\begin{table}[h]
	\caption{Background Period Definition and Description}
	\label{tbl2_definition}
	\centering
	\begin{tabular}{@{}c|l@{}}
		\toprule
		\textbf{Background Period} & \textbf{Description} \\ \midrule
		Quiet/Night ($ NP $) & 
		\begin{tabular}[c]{@{}l@{}} There is minimal or no human outdoor\\ activity going on.\end{tabular}\\
		Active ($ AP $) & 
		\begin{tabular}[c]{@{}l@{}} There is some outdoor activities going \\ on.
		\end{tabular}\\
		Rush Hour period ($ RP $) &
		\begin{tabular}[c]{@{}l@{}} There is large number of outdoor activities \\ going on.
		\end{tabular}\\   \bottomrule
	\end{tabular}
\end{table}

\begin{figure}[h]
	\centering
	\includegraphics[width=0.8\textwidth]{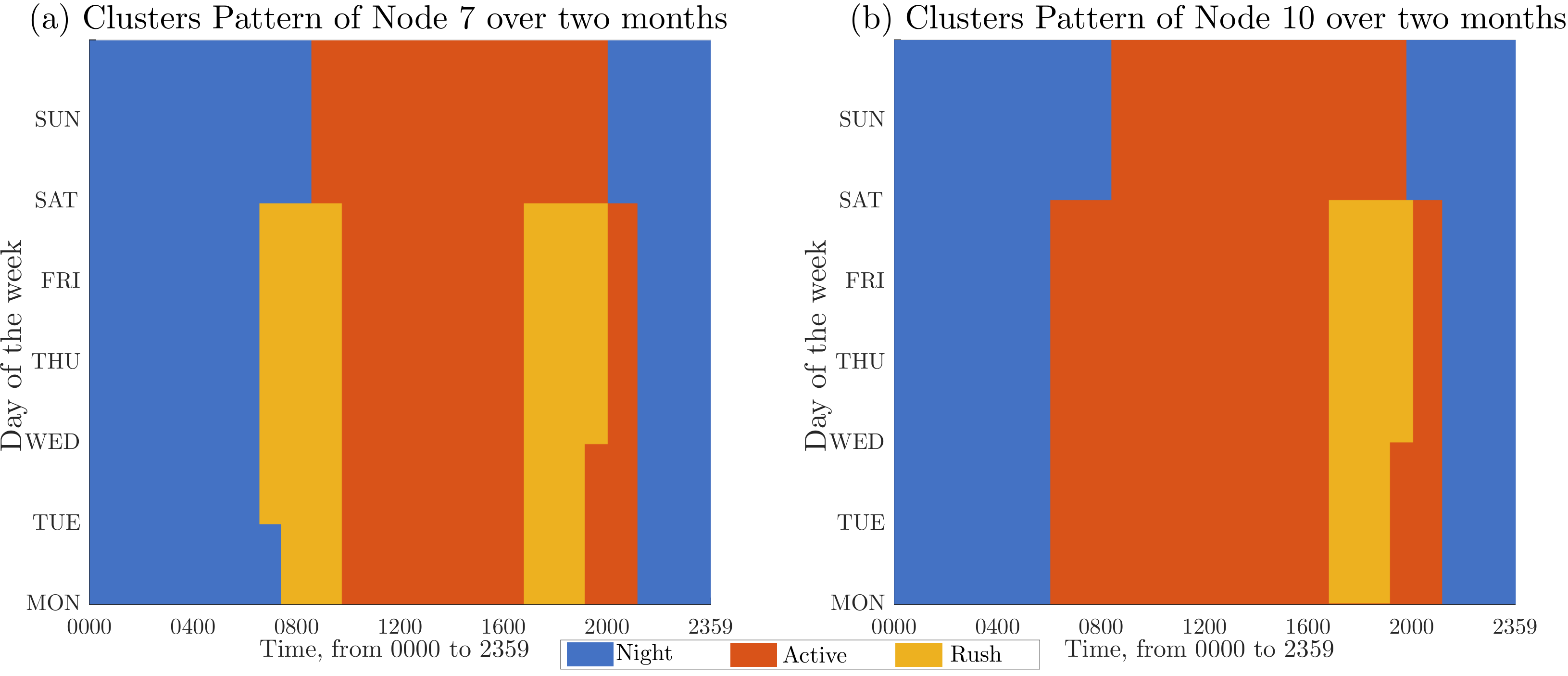} 
	\caption{Average of periods pattern over a two months period ($1$st July 2016 - $31$st August 2016) for Node $7$ and $10$.}
	\label{fig8_avgPattern}
\end{figure}

After defining individual background noise periods, we observed some background noise behavior from the data shown in Fig.~\ref{fig8_avgPattern}. 
For instance, there is long $NP$ on weekend, which is possible due to the majority of the residents are not working at weekend. 
This also contributed to a shorter $RP$ at weekend.
In addition, there is an increment of the outdoor activities (scattered points) on Sunday compared to Saturday and weekday. 
In the Fig.~\ref{fig8_avgPattern}(a), there is two obvious rush hour periods during morning and evening, where in Fig.~\ref{fig8_avgPattern}(b), it only shows the evening rush hour period.
The reason why sensor node $7$ have two peaks detected is because it is located in the proximity of road, railway line, and pedestrian path. 
Meanwhile, sensor node $10$ is located far away from these locales (roughly $400$m apart) and capture less background noise compared to sensor node $7$.

However, if we attempt to detect human outdoor activity only with the noise level, we are not able to differentiate the background noise and human outdoor activity.
Instead, we try to understand the background noise across different time of the day and week to provide a noise pattern study before detecting any outdoor activity.

\subsection{Outdoor Activity Detection}
\label{OAD}

After identifying the background period for each sensor, we need to further investigate methods to distinguish an outdoor activity and background noise using outlier detection. 
Outdoor activity can be defined as the presence of human outdoor activity in a particular PoI. 
This includes kid's playing and large group chatting, which happens in the vicinity of the sensor node.
As observed, the average value level of $RP$ is much higher compared with $AP$ and $NP$ with different deviation values.
In Fig.~\ref{fig7_backgroundAvgNoise}, we notice that the average noise value varies through different times of the day, and it is hard to draw a meaningful conclusion out of it.
Therefore, a simple and robust method such as applying a threshold is not suitable under these constraints.
In addition, the variation of background noise collected in each sound sensor is quite high, requiring different thresholding values and it is not efficient to manually devise a threshold for each sound sensor.
The chances of This further increase of false-positive (falsely label background noise as outdoor activity) and false negative (falsely label outdoor activity as background noise) in the outdoor detection model.
Therefore, data samples from different background noise periods need to be taken into consideration when identifying the outdoor activities separately. 

In the outdoor activity detection module, we calculate a chi-square statistic for each sound histograms samples to distinguish background noise and human outdoor activity.
Chi-square distribution with $k$ degrees of freedom is described as a distribution of the squares for $k$ independent standard normal random variables. 
The corresponding of the Chi-square distribution with $k$ degrees of freedom can be calculated when optimal $k$ PCs are achieved through independence between PCs.
For example, we can show two PCs are independent by proving the product of two eigenvectors is zero by applying the symmertricity of two covariances as follows:

Suppose $ \lambda_{i}\neq\lambda_{j} $ and $ \Sigma e_{i}=\lambda_{j} e_{i} $ and $ \Sigma e_{j}=\lambda_{j} e_{j}$ where $ \lambda_{i} $, $ \lambda_{j} $ are eigenvectors of the symmetric matrix $ \Sigma $, with corresponding eigenvectors, $  e_{i} $ and $  e_{j} $ respectively. 
Then We have the following:
\begin{align*}
(\lambda_{i} - \lambda_{j})( e_{i}-e_{j})=&(\lambda_{i}e_{i},e_{j})-(e_{i},\lambda_{j}e_{j})\\
=& (\Sigma e_{i},e_{j})-(e_{i},\Sigma e_{j})
\end{align*}

By using the symmetricity of $\Sigma $ we can rewrite the equation as below:
\begin{align*}
(\Sigma e_{i},e_{j})-(e_{i},\Sigma e_{j}) =& \Sigma e_{i}.e_{j}-e_{i}.\Sigma e_{j} \\
=& e_{i} \Sigma^{T}e_{j}-e_{i}.\Sigma e_{j} \\
=& e_{i} \Sigma e_{j}-e_{i}.\Sigma e_{j} = 0
\end{align*}
Since $ \lambda_{i}\neq\lambda_{j} $, it shows that product of any two eigenvectors of a symmetric matrix is zero.

After denoting zero symmetric matrix and optimality of $k$ PCs, we compute the corresponding chi-square statistic ${\chi}^2_{(i)}$ of $k$ degrees of freedom for sound histogram  samples as follows:
\begin{equation}
{\chi}^2_{(i)} = pc^2_{(1,i)}+pc^2\textbf{}_{(2,i)}+...+pc^2_{(k,i)},
\end{equation}
where $pc^2_{(n,i)}$ is the standardized \textit{PC} score of the $i^{th}$ sound histogram (all sound histograms correspond to the same background noise period) and $n \in {\{1,...,k\}}$.

If the background noise remains stationary (sound histograms within the same cluster yields similar property) between sound histograms corresponding to the same time of the day, we can denote that particular data sample has a lower probability of being classified as outdoor activities.
The intuition behind is that a sound histogram, which possesses lower probability values represents the outdoor activities that deviate significantly from a background noise period. 
Using the aforementioned concept, we can further formulate a binary output for activity detection using the following equation:
\begin{equation}
h_{activity} = \left\{ 
\begin{matrix}
1 &\text{ if } & \chi _{(i)}^2 \ge \beta  \hfill \cr 
0 &\text{ if } & \chi _{(i)}^2 < \beta  \hfill \cr \end{matrix} \right.
\label{eqn:soundsUtz}
\end{equation}
where $\beta$ is a critical value in $k$ degrees of freedom chi-square distribution and $h_{activity}$ is a binary value that represents the outdoor activity for a given sound histogram samples.
In addition, $\chi _{(i)}^2 $ denotes the activity level for each sound histogram sample.
As a remark of the outdoor activity detection, the period detected as raining would be labeled as zero outdoor activities, so that both events do not overlap with each other.

Next, we need to determine the value of $\beta$, which it is required to define the lower probability separation of the sound histograms sample into outdoor activity and background noise.
We devise the approach of using quantile-quantile ($q$-$q$) plot~\cite{gibbons2014nonparametric,Widhiarso2016Examining,Leys2018Detecting} of a background noise period to visually identify the deviation sound sample at a given day of study.
In the $q$-$q$ plot, computed chi-square quantiles are plotted against the theoretical chi-square quantiles, and we can assess whether our computed values plausibly come from the same theoretical distribution.
If the value does not fall in the same theoretical distribution, it implies that particular interval has a higher probability of outdoor activity.

\begin{figure}
	\centering 
	\includegraphics[width=0.95\textwidth]{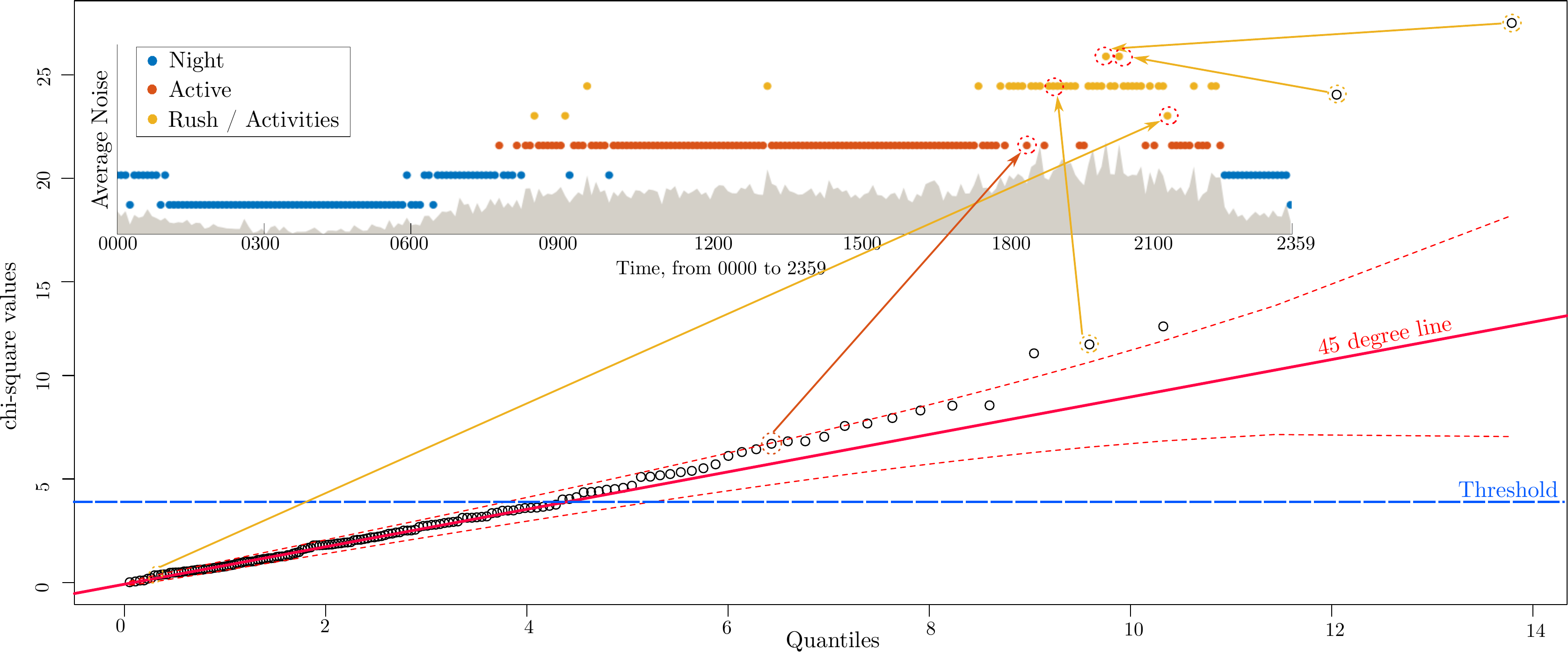}
	\caption{Quantile-Quantile plot of the chi-square distribution for the active period on Saturday $16$th July $2016$ (Node $7$)}	 
	\label{fig9_qqPlot}
\end{figure}

We chose a Saturday ($16$th July $2016$) as one of the days to study outdoor activity using sensor node $7$. 
The $q$-$q$ plot of the noise samples is illustrated in Fig.~\ref{fig9_qqPlot}.
From the $q$-$q$ plot, we observe chi-square values begin to deviate significantly from the $\angle{45}$ line beyond a certain threshold $T$. 
Based on the patterns, we observe some sound samples are related to a different distribution compared to a chi-square distribution.
For instance, the period of $18$:$00$ to $21$:$00$ is supposed to be a rush period, but one point that is highlighted by the orange arrow shows an active period instead.
These data point that falls outside the red dotted line, it is considered as outliers and it has a higher probability of an outdoor activity being detected.
Meanwhile, those located within the red dotted line or close to $45\deg$ line is either background noise or lower probabilities labeled as outdoor activities.
By tracing back the sound histograms temporal element, we can obtain the occurrence time for data samples and reconstruct the time series data with a label for each sound histogram samples.
Note that, the same procedure can be repeated for other sensor nodes and obtain a similar result.

Apart from the $q$-$q$ plot approach, $T$ also can be estimated from other change point estimation methods such as the hybrid approach~\cite{Boutoille2010hybrid}, Bayesian change point estimation~\cite{Aminikhanghahi2017survey}, and wavelet methods \cite{Ogden1996Change}, etc.
Nonetheless, we use the $T=0.43$ as $\beta$ for the outdoor detection method for our system model.
Thus, this concludes the outdoor activity detection module, and we will perform validation in the next section.

\section{Evaluation} 
\label{experi}

\begin{figure}[h]
	\centering 
	\includegraphics[width=1\textwidth]{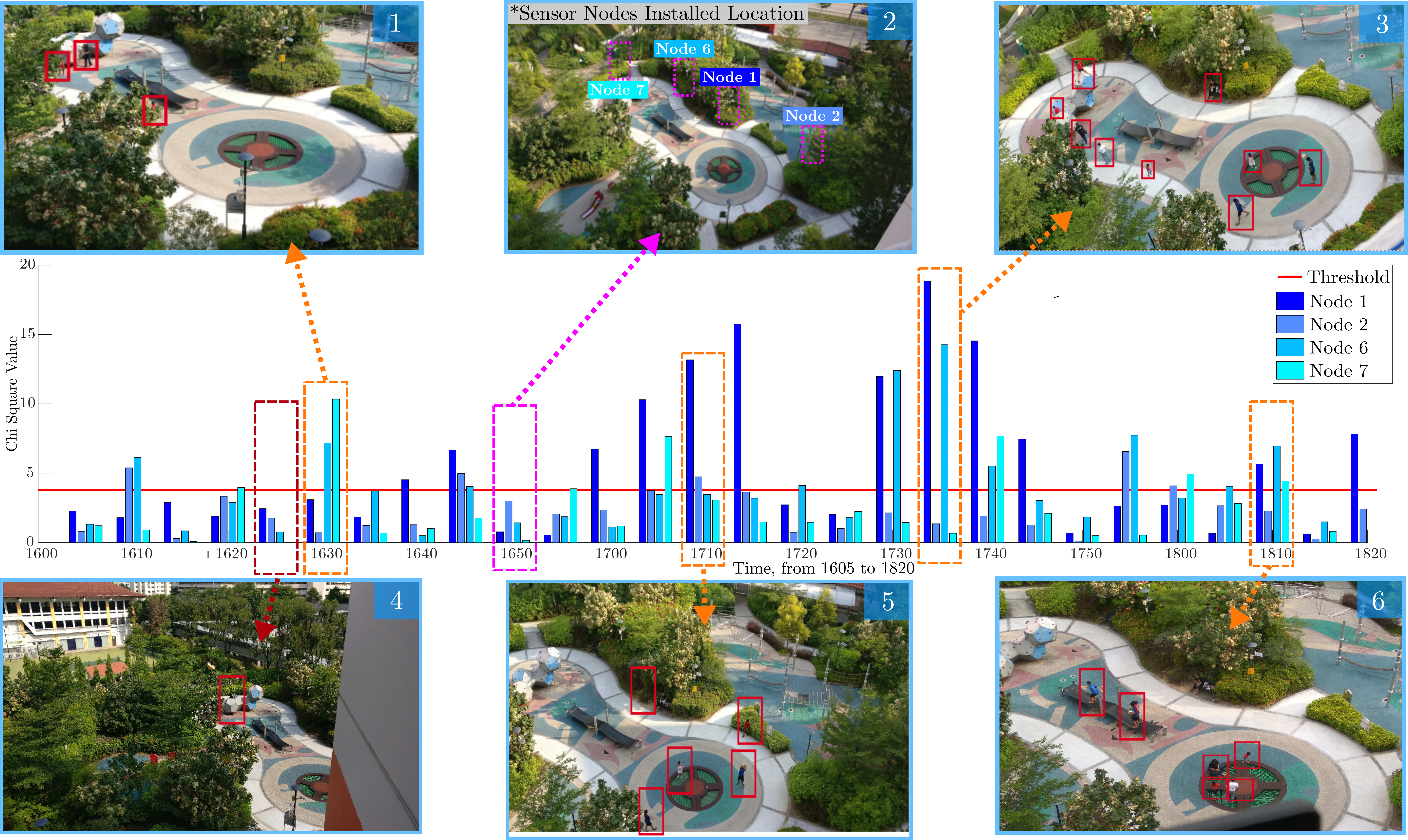}
	\caption{Obtained chi-square values with the ground truth images (snipped from the recorded video) for each sensor nodes on $20$th July $2017$ evening (only four out of seven sensor nodes are presented here). }	 
	\label{fig10_validation}
\end{figure}

In this section, we validate the detected events by the proposed method against the ground truth to verify the accuracy of the proposed outdoor activity detection model.
The same open space playground located near the residential area studied in the previous section is also used to collect ground-truth data.
There are a total of $7$ sensor nodes installed throughout the playground area ($1650$m$^2$), which is also next to the metropolitan railway line, bus-stop, and main road.
These combinations of nearby amenities produce a PoI with noisy ambient noise varying from time to time, which is an ideal test case to prove the effectiveness of our model.
We collect the ground-truth data on a Saturday evening ($20$th July $2017$) for $5$ hours starting from $14$:$00$ to $19$:$00$, where there is a high probability of human outdoor activities.
Although there are $7$ sensor nodes located at the PoI, we only focus on $4$ sensor nodes around the premises that are within our range of sight as shown in Fig.~\ref{fig10_validation}(2) with the magenta arrow. 
We captured the video for $5$ hours and invited $3$ volunteers to label the outdoor activity time stamp in the video as ground-truth data.
Throughout the data collection, there are roughly two hours ($16$:$00$ - $18$:$20$) of activities being recorded as shown in Fig.~\ref{fig10_validation}.

Using a similar threshold from the previous section, we identified the outdoor activity and compared them to the ground-truth.
The aggregation of $4$ sensors is achieved through a ``or'' function, where one sensor detected outdoor activity represents the presence of human outdoor activity in that PoI at that particular timestamp.
After comparing the computed result and ground-truth data, it is not hard to see that the number of people in each image closely resembles the aggregated output of the sensor nodes.

\begin{table}[h]
	\caption{Experimental Results}
	\vspace{-0.4cm}
	\label{tbl_expResult}
	\begin{center}
		\begin{tabular}{@{}l|c|c|c@{}}
			\toprule
			Methods & \begin{tabular}[c]{@{}c@{}}True \\ Detected\end{tabular} & \begin{tabular}[c]{@{}c@{}}False \\ Positive\end{tabular} & \begin{tabular}[c]{@{}c@{}}False\\ Negative\end{tabular} \\ \midrule
			Raw data & 	 65.0\% & 	 35.0\% & 	0.0\% \\
			Raw data + WT & 	60.2\%& 	34.3\% & 	5.5\% \\
			Raw data + PCA & 	50.0\% & 	33.5\% & 	16.5\% \\
			Raw data + PCA + WT & 85.8\% & 3.5\% & 10.7\% \\ \bottomrule
		\end{tabular}
	\end{center}
\end{table}

Next, we generate the confusion matrix based on the comparison of different processing methods and ground-truth.
The result is shown in Table~\ref{tbl_expResult}.
It can be observed that approaches without applying WT and PCA (raw data only, raw data and PCA as well as raw data and WT) yields higher false positive, which contributes to lower true detected values (65.0\%, 60.2\%, and 50\% respectively).
To understand the reason behind false-negative rates of $10.7$\% for proposed approach using PCA and WT, we study the ground-truth at the corresponding timestamps. 
It reveals that these false-negative events corresponded when there are less than one or two people within the PoI that is quiet.
An example of the false-negative event can be found in Fig.~\ref{fig10_validation}(4).
It shows there is only one person in the PoI at that time-stamp and remains quiet and hence not adequate sound information to capture the event accurately.
As for other occupancy detection, it remains quite accurate most of the time. 
Thus, we can draw a conclusion that the system model we proposed work effectively and requires less human effort to provide training data.

\section{Conclusion} 
\label{concl}
In this paper, we have shown how the sound histogram can be used as a data reduction method to collect insights about the residential area rather than sending data samples frequently.
Using methods such as WT and PCA, we demonstrate it is possible to generate more robust and meaningful feature sets before performing a clustering algorithm to group similar data patterns.
In order to further extract insights from the sound histogram data, we devise a system model to detect background noise period, human outdoor activity, and raining period.
Lastly, we validate our model against ground-truth data and managed to identify human outdoor activity with an accuracy of $85.5$\% using only sound sensor.

In future works, we expect the system model to adapt different sources of data to enrich the outdoor activity detection patterns. 
Multiple sensors will be used as a foundation of the sensor fusion to generate more accurate data is also part of the study.
Moreover, the trade-off between normal sampling rate versus sound histograms could be further investigated from a hardware or information theory perspective is quite interesting. 
Also, the optimal histogram bin size and range choices also can be throughout investigated in order to increase detection accuracy.
This will provide us a more optimal system to design data collection platforms from different aspects such as energy consumption of sensor nodes, communication overhead, and data quality.

\section*{Acknowledgment}
This work is supported by Singapore Ministry of National Development (MND) Sustainable Urban Living Program, under grant no. SUL2013-5, ``Live-able Places: A Building Environment Modeling Approach for Dynamic Place Making” project, and especially appreciate the useful discussion and help from the collaborators from MND, HDB, and URA.

\bibliography{bibSpace}

\end{document}